\documentclass[conference]{IEEEtran}

\ifCLASSINFOpdf
\else
\fi
\usepackage[T1]{fontenc}
\usepackage{textcomp}
\usepackage{graphicx}
\usepackage{amsmath}
\usepackage{booktabs}
\usepackage{algorithm}
\usepackage{algpseudocode}
\usepackage{tikz}
\begin{document}

\title{Multilayer Forensic Tampering Detection

}

\author{\IEEEauthorblockN{ Titouan Millet, Thomas Valade, François Gonnet and Mounira Msahli}
\IEEEauthorblockA{Télécom Paris, Institut Polytechnique de Paris}
\IEEEauthorblockA{(titouan.millet, thomas.valade, mounira.msahli)}@telecom-paris.fr and francois@ubg-interactive.fr\\}

\maketitle

\begin{abstract}
With the proliferation of free online editing tools, altering or forging pdf documents has become trivially easy, often leaving no visual traces on screen. This paper introduces a two-stage forensic pipeline. An initial security-gating layer validates format compliance and flags embedded malicious payloads and a forensic engine that inspects internal objects across metadata, visual overlays, and dual-source OCR consistency, etc... A weighted scoring engine aggregates these forensic indicators into an interpretable risk score is proposed. The approach was validated on real medical work-stoppage certificates.
\end{abstract}

\IEEEpeerreviewmaketitle

\section{Introduction}
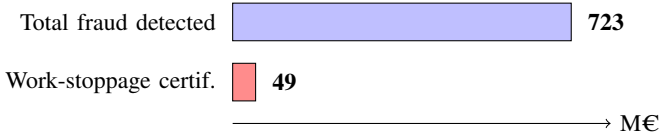
\begin{figure}[htp]
\centering
\begin{tikzpicture}[font=\small]
    \draw[->] (0,0) -- (5.0,0) node[right]{M\texteuro{}};
    \draw[fill=blue!25] (0,1.1) rectangle ++({723*0.0062},0.5);
    \node[anchor=east] at (-0.1,1.35) {Total fraud detected};
    \node[anchor=west] at ({723*0.0062+0.1},1.35) {\textbf{723}};
    \draw[fill=red!45] (0,0.3) rectangle ++({49*0.0062},0.5);
    \node[anchor=east] at (-0.1,0.55) {Work-stoppage certif.};
    \node[anchor=west] at ({49*0.0062+0.1},0.55) {\textbf{49}};
\end{tikzpicture}
\caption{Assurance Maladie fraud detected and stopped in 2025: total vs. falsified work-stoppage certificates alone \cite{7}}
\label{fig:fraud}
\end{figure}

The digitization of administrative exchanges has made PDF the standard for transmitting official documents, from invoices to identity documents and medical certificates. Yet PDF was engineered for visual fidelity, not cryptographic immutability: free online editing tools now make targeted tampering trivial to perform while leaving the rdered page perfectly seamless. The stakes are concrete: in France alone, the national health insurer reported 723 million euros of fraud detected and stopped in 2025, of which 49 million euros were tied to falsified work-stoppage certificates alone \cite{7} (Fig.~\ref{fig:fraud}). Manual visual inspection can no longer keep pace, creating an urgent need for automated, objective detection tools. \\

Existing approaches rely mainly on basic metadata inspection, structural analysis of incremental updates, or opaque commercial platforms that require uploading sensitive files to third-party infrastructure \cite{1} limitations our module-level, locally-run, document-aware pipeline is designed to address. We present a modular application whose primary goal is to facilitate and automate the detection of PDF tampering or compromise, built on eight independent forensic modules (format validation, malware detection, metadata, structure, fonts, images, visual overlay, and dual-source OCR) aggregated into a weighted, interpretable risk score. The rest of this paper details the pipeline architecture, each inspection module, and results on real-world cases. 

\section{Proposed Analysis pipeline}
As indicated in algorithm 1 below, every analysis function is wrapped by a common decorator provides end-to-end processing integrity and a verifiable audit trail, guaranteeing that the inspection procedure neither mutates the original sample nor obscures diagnostic steps. \par
\begin{algorithm}[htp]
\caption{\texttt{AnalysisEngine.run(pdf\_bytes)}}
\label{alg:pipeline}
\begin{algorithmic}[1]
\footnotesize
\If{\textbf{not} pdf\_validator.is\_valid(pdf\_bytes)}
    \State \Return Score(100, \textsc{Invalid}) \Comment{Stage 1 halt}
\EndIf
\If{malware.scan(pdf\_bytes).detected}
    \State \Return Score(100, \textsc{Critical: Host Threat}) \Comment{quarantine}
\EndIf
\State $meta \gets$ metadata.analyze(pdf\_bytes)
\If{$meta$.creator\_producer\_risk = \textsc{High}}
    \State \Return Score(100, \textsc{Critical: Forgery}) \Comment{Stage 2}
\EndIf
\State $R \gets \{$fonts, structure, images$\}$.analyze(pdf\_bytes)
\If{$R$.structure.has\_drawings \textbf{or} $R$.images.count $>0$}
    \State $R$.coverage $\gets$ coverage.analyze(pdf\_bytes)
\EndIf
\If{$R$.images.count $>0$}
    \State $R$.ocr $\gets$ ocr.analyze(pdf\_bytes)
\EndIf
\State $score \gets$ scoring.compute($R$) \Comment{Eq.~\eqref{eq:scoring}}
\State \Return $score$, analyze.report($R$, $score$)
\end{algorithmic}
\end{algorithm}

The pipeline strictly decouples host protection from forensic forgery analysis across two successive phases (Algorithm~\ref{alg:pipeline}):
\begin{itemize}
    \item \textbf{Ingestion and Security Gating}: An upfront sanitization gate ensures that the uploaded binary complies with basic PDF specifications and does not serve as a payload targeting the host environment. An invalid file format or an active exploit halts execution immediately, neutralizing risks to the infrastructure before deeper parsing begins.
    \item \textbf{Forensic Forgery Inspection}: Cleared documents undergo multi-layer forensic scrutiny. An eliminatory check on high-risk Creator/Producer signatures immediately assigns a maximum risk score (100\%), while remaining modules evaluate structural and visual consistency to uncover fraudulent modifications.
\end{itemize} \par

 For each forensic module or filter, we consider $m \in \mathcal{A}$ (weight $w_m$, maximum attainable raw score $M_m$), the triggered flags $f \in \mathcal{F}_m$ contribute a capped ratio $r_m$; the final score normalizes over active modules only, so a page naturally lacking optional features (e.g. no raster imagery) is never penalized for a module that had nothing to analyze:
\begin{align}
r_m &= \frac{\min\!\left(100,\ \textstyle\sum_{f \in \mathcal{F}_m} s_f\right)}{M_m} \label{eq:ratio}\\
\text{score} &= \text{round}\!\left(100 \times \frac{\textstyle\sum_{m} r_m\, w_m}{\textstyle\sum_{m} w_m}\right) \label{eq:scoring}
\end{align}
with weights $w_m \in \{25,25,20,15,15\}\%$ for \{metadata, structure, coverage, fonts, images\}. The resulting score maps to an interpretable risk level (Table~\ref{tab:levels}), with any eliminatory flag short-circuiting directly to the maximal \textsc{Critical} level.

\begin{table}[htp]
\centering
\caption{Risk level mapping}
\label{tab:levels}
\footnotesize
\begin{tabular}{@{}lccl@{}}
\toprule
Score range & & Risk level & \\
\midrule
0 & & Compliant & (\emph{Conforme}) \\
1--15 & & Low & (\emph{Faible}) \\
16--35 & & Moderate & (\emph{Modéré}) \\
36--60 & & High & (\emph{Élevé}) \\
61--99, or 100 (elim.) & & Critical & (\emph{Critique}) \\
\bottomrule
\end{tabular}
\end{table}

\section{Implementation and results}


The pipeline was exercised on a working corpus of real medical work-stoppage certificates, deliberately falsified variants, and known malicious PDFs. Table~\ref{tab:cases} summarizes three representative cases spanning the full risk spectrum; the two most evidential are illustrated below.

\begin{table}[htp]
\centering
\caption{Representative cases across the risk spectrum}
\label{tab:cases}
\footnotesize
\begin{tabular}{@{}lcc@{}}
\toprule
Case & Score & Level \\
\midrule
Falsified certificate & 54\% & High \\
Malicious payload & 100\% (elim.) & Critical \\
Unmodified letter & 33\% & Moderate \\
\bottomrule
\end{tabular}
\end{table}

\begin{figure}[htp]
    \centering
    \includegraphics[width=0.55\linewidth]{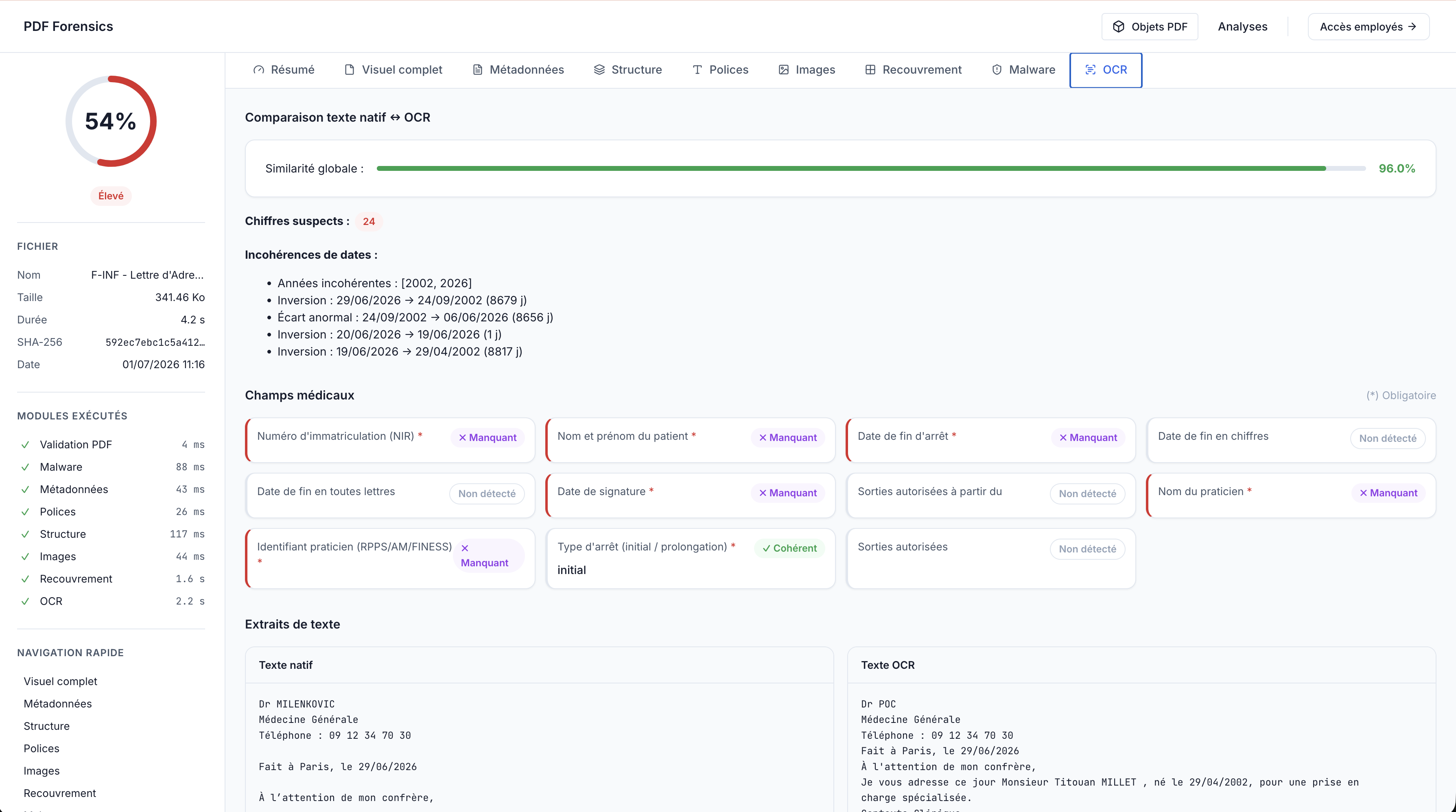}
    \caption{Native content stream vs. rendered (OCR) text on the same document, exposing a substituted name invisible on screen}
    \label{fig:ocr}
\end{figure}

\textbf{Structural and metadata detection} (Fig.~\ref{fig:ocr}): as an example, on the falsified certificate, structure and coverage flagged a 100\% page stacking rate, and the native content stream read \emph{“Dr MILENKOVIC”} where the rendered page displayed \emph{“Dr POC”}  invisible to a human reader relying on the rendered document alone. The same file carried 24 suspicious digits, 5 date inconsistencies, and an \texttt{AppendMode} Producer marker absent from XMP. \par

\begin{figure}[htp]
    \centering
    \includegraphics[width=0.55\linewidth]{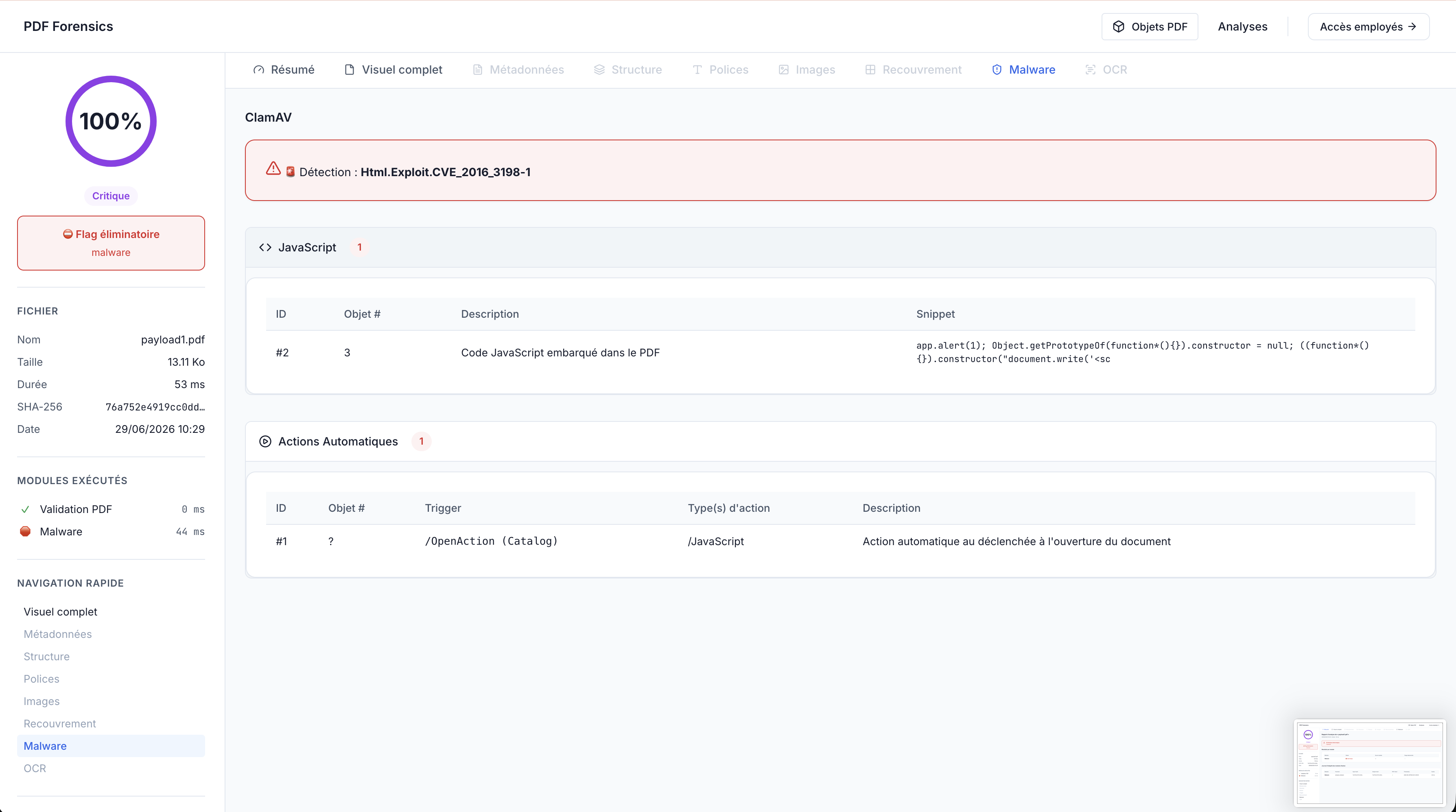}
    \caption{Eliminatory malware detection: embedded JavaScript triggered at document opening}
    \label{fig:malware}
\end{figure}

\textbf{Infrastructure protection} (Fig.~\ref{fig:malware}): the eliminatory malware check quarantined the payload --- ClamAV \texttt{Html.Exploit.CVE\_2016\_3198-1}, JavaScript triggered via \texttt{/OpenAction} within 44~ms, before any forensic module ran, rather than evaluating document authenticity. \par

\textbf{Limitations and future work}: the module weighting is not immune to false positives (Table~\ref{tab:cases}, row~3): a visually unmodified letter was scored \emph{Moderate} primarily because it carried no XMP stream, a field many legitimate PDF generators simply never write. Any single metadata-absence flag is thus a weak signal in isolation, which is precisely why the scoring engine combines eight independent modules rather than triggering on one flag alone; reducing this residual false-positive rate on documents with sparse metadata remains open work. The proposed approach is extensible, allowing additional filtering and forensic analysis modules to be defined and integrated as new fraud or tampering patterns. Another direction for future improvement would be to incorporate AI-based approaches to improve detection accuracy and overall system effectiveness through data-driven learning. 

\section{Conclusion}
This work presented a forensic pipeline designed to facilitate the detection of PDF tampering. Results show that cross-referencing the internal structure of PDF files with their rendered content can reveal alterations that are not apparent during simple visual inspection or analysis limited to metadata alone. Several future improvements are proposed. 



\end{document}